\documentclass[aps,twocolumn,superscriptaddress,preprintnumbers,showpacs,tightenlines]{revtex4}
\usepackage{amssymb}
\usepackage{epsfig,graphicx,times}

\newcommand{\beq}{\begin{equation}}
\newcommand{\eeq}{\end{equation}}

\newcommand{\bea}{\begin{eqnarray}}
\newcommand{\eea}{\end{eqnarray}}

\begin{document}

\preprint{Second Draft}
\title{Quantum effects in thermal conduction: Nonequilibrium quantum discord and entanglement}

\author{Lian-Ao Wu}
\affiliation{Department of Theoretical Physics and History of Science, The Basque Country
University (EHU/UPV) and IKERBASQUE -- Basque Foundation for Science, 48011,
Bilbao,Spain}

\author{Dvira Segal}
\affiliation{Chemical Physics Group, Department of Chemistry and Center for Quantum
Information and Quantum Control, University of Toronto, 80 St. George
street, Toronto, Ontario, M5S 3H6, Canada}

\begin{abstract}
We study the process of heat transfer through an entangled pair of two-level
system, demonstrating the role of quantum correlations in
this nonequilibrium process. While quantum correlations generally degrade with increasing
the temperature bias, introducing spatial asymmetry leads to an intricate behavior:
Connecting the qubits unequally to the reservoirs one finds that quantum correlations
persist and {\it increase} with the temperature bias when the system is more weakly linked to the hot reservoir.
In the reversed case, linking the system more strongly to the hot bath, the opposite, more natural
behavior is observed, with quantum correlations being strongly suppressed
upon increasing the temperature bias.
\end{abstract}

\maketitle

\section{Introduction}

Understanding thermal energy transfer at the nanoscale has recently become a topic
of great interest in nanotechnology \cite{Dubi}, with proposals for new
devices that can actively control heat conduction and information
storage: thermal rectifiers \cite{Rectifier},
thermal logic operators \cite{BaowenL}, and  memory devices \cite{BaowenM}.
As some of these devices have already been realized \cite{RectifierE, MemoryE},
one should note that these thermal elements typically have been analyzed within the principles
of classical mechanics, operating at room temperature.
Assessing the role of quantum effects in the operation of such systems is of
fundamental and practical importance, for building quantum
devices fighting relaxation and decoherence processes,  operating under nonequilibrium conditions.

The typical setup of interest in this context includes a small system, with few degrees of freedom
(nanobeam, linear molecule, spin chain),
connected at its ends to two large reservoirs (solids, metals),
maintained at different temperatures. In the steady state limit
a constant heat current flows through the system.
The thermal conductance of such a nanoscale junction has been primarily
simulated using either of the following three
approaches: (i) The Landauer's formula \cite{Kirczenow} combined with first-principles calculations
of the Hamiltonian force constants \cite{NanoL}, (ii)
the non-equilibrium Green's function method \cite{Green,Tu}, or (iii)
classical molecular dynamics simulations \cite{Classical}.
The Boltzmann-Peierls phonon transport theory \cite{BP,Spohn},
mixed classical-quantum simulations \cite{Mixed}, and quantum master equation methods
\cite{Master,SegalM,WuRectif} are other methods developed for predicting the
conductance properties of different objects.
For systems with few degrees of freedom the latter method is of particular
interest, as the kinetic equations of motion can be
derived, under some approximations, from the fundamental quantum equations of
motion, as explained below.
For simple model systems these equations can be analytically solved, providing
insight on the microscopic dynamics \cite{SegalM}.

Entanglement and quantum discord \cite{Qcorr} are quantum correlations
with no classical counterpart
which can be used as tools for identifying and distinguishing the quantum
aspects in the thermal transport process from the classical ones \cite{EntagT1,EntagT2,EntagT3}.
It is our objective here to consider a simple (yet involving many-body interactions)
thermal conducting junction, to study its transport characteristics within the quantum master equation method, and to
evaluate the role of quantum effects in the energy transport process through the entanglement
and discord measures.
In particular, as the steady state concurrence in quantum open systems  has been already
analyzed  \cite{EntagT1,EntagT2,EntagT3}, it is of interest 
to explore its relation to the steady state discord measure, quantifying non-classical correlations
beyond entanglement.



The particular system examined here includes a pair of a two-level
system under a magnetic field, placed in between two thermal reservoirs. We use this
model as a case study for illustrating the intricate role of the nonequilibrium
condition on quantum correlations in the system.
We analytically calculate the amount of entanglement and discord in this model
at different baths temperatures, 
demonstrating that even at relatively high temperatures
quantum correlations play a role in the energy transfer process.
Furthermore, by introducing asymmetry, we show that, counter-intuitively,
quantum correlations may be {\it enhanced} upon increasing the
temperature bias across the system, depending on the bias polarity.


\section{Model}

Our model includes a quantum system coupled to two different thermal reservoirs.
The system incorporates two interacting qubits in a magnetic field
subjected to the $XY$ interaction.
These qubits, 1 and 2, are separately coupled to independent reservoirs $H_L$ and $H_R$,
respectively,
maintained in thermal equilibrium at temperatures $T_{\nu}$, $\nu=L,R$.
The total Hamiltonian is given by
\bea
H=H_S+\sum_{\nu}H_{\nu}+V_L+V_R,
\eea
where the two-qubit Hamiltonian is
\bea
H_{S}=\frac{\epsilon }{2}(\sigma _{1}^{z}+\sigma _{2}^{z})+\frac{\kappa }{2}%
(\sigma _{1}^{x}\sigma _{2}^{x}+\sigma _{1}^{y}\sigma _{2}^{y}).
\eea
Here $\sigma^{i}$ ($i=x,y,z$) are the Pauli matrices.
The system Hamiltonian can be diagonalized to produce the "system diagonal" basis
with four states $n=1,2,3,4$,
\bea
\left\vert 1\right\rangle &=&\frac{1}{\sqrt{2}}(\left\vert \downarrow
\uparrow \right\rangle -\left\vert \uparrow \downarrow \right\rangle
), \,\,\,\ E_{1}=-\kappa
\nonumber\\
\left\vert 2\right\rangle &=&\left\vert \downarrow \downarrow \right\rangle
, \,\,\,\ E_{2}=-\epsilon
\nonumber\\
\left\vert 3\right\rangle &=&\left\vert \uparrow \uparrow \right\rangle, \,\,\,\
E_{3}=\epsilon
\nonumber\\
\left\vert 4\right\rangle &=&\frac{1}{\sqrt{2}}(\left\vert \downarrow
\uparrow \right\rangle +\left\vert \uparrow \downarrow \right\rangle
), \,\,\,\ E_{4}=\kappa.
\label{eq:basis}
\eea
The states are ordered assuming that the inter-spin interaction is strong,
$\kappa >\epsilon>0$.
As the qubits are identical, the symmetric conditions
$\omega_{43}=\omega_{21}$ and $\omega_{42}=\omega_{31}$ follow, with $\omega_{nm}=E_n-E_m$.
We assume that system-bath interactions are separable, and use the following bipartite form,
\bea
V_{\nu}=B_{\nu}S_{\nu},
\eea
with $B_{\nu}$ as a bath operator and $S_{\nu}$ a system operator. Specifically, we use
$S_L=\sigma_1^x$ and $S_R=\sigma_2^x$.
In the basis of Eq. (\ref{eq:basis}) these operators translate to
\bea
S^{L}&=&\frac{1}{\sqrt{2}}(-|1\rangle \langle 2|+|1\rangle \langle
3|+|2\rangle \langle 4|+|3\rangle \langle 4|+h.c.)
\nonumber\\
S^{R}&=&\frac{1}{\sqrt{2}}(|1\rangle \langle 2|-|1\rangle \langle
3|+|2\rangle \langle 4|+|3\rangle \langle 4|+h.c.).
\label{eq:S}
\eea
We next calculate the heat current and the quantum entanglement and discord.
Note that we have not yet specified a  model for the  reservoirs,
as the calculations can be formally done for different bath realizations \cite{WuRectif}.

\section{Steady state dynamics}

Under nonequilibrium conditions, $T_L\neq T_R$, in the long time limit,
we present next the following quantities: (i) the system's population,
(ii) the steady state heat current, and (iii) the resulting quantum correlations.

\subsection{Levels' population}
We follow standard weak coupling schemes \cite{Breuer} and use the Born-Markov approximation.
Beginning with the Liouville equation, the method first involves the assumption of
weak system-bath interactions. Furthermore, we apply the Markovian limit,
assuming that the reservoirs' characteristic timescales are shorter than the
subsystem relaxation time. Under these approximations a master equation
for the states population $P_n$ ($n=1,2,3,4$) can be readily obtained,
\bea
\dot P_n(t)&=&\sum_{\nu,m}|S_{mn}^{\nu}|^2P_m(t) k_{m \rightarrow n}^{\nu}
\nonumber\\
&-&P_n(t) \sum_{\nu,m} |S_{mn}^{\nu}|^2 k_{n\rightarrow m}^{\nu}.
\label{eq:master}
\eea
Details about this derivation (for the two-bath case) are given in Refs. \cite{SegalM}
and \cite{WuRectif}.
Here $S^{\nu}=\sum S_{mn}^{\nu}|m\rangle \langle n|$.
The Fermi golden rule transition rates are given by
\bea
k_{n\rightarrow m}^{\nu}=\int_{-\infty }^{\infty
}d\tau e^{i\omega_{nm}\tau }\left\langle B_{\nu}(\tau
)B_{\nu}(0)\right\rangle_{T_{\nu}},
\label{eq:rate}
\eea
where $B_{\nu}(\tau )=e^{iH_{\nu}\tau }B_{\nu}e^{-iH_{\nu}\tau }$
are interaction picture operators, The thermal average is given by
$\langle O \rangle_{T_{\nu}}={\rm Tr} [e^{-H_{\nu}/T_{\nu}}O]/{\rm
Tr} [e^{-H_{\nu}/T_{\nu}}]$. Note that the rates are evaluated at a
specific subsystem frequency. For example, in Eq. (\ref{eq:rate})
the relevant energy scale is $\omega_{nm}=E_n-E_m$. Solving Eq.
(\ref{eq:master}) in steady state we obtain the population
\bea
P_{1} &=&\frac{W_{12}W_{13}}{\left( W_{12}+W_{21}\right) \left(
W_{13}+W_{31}\right) }
\nonumber\\
P_{2} &=&\frac{W_{21}W_{13}}{\left( W_{12}+W_{21}\right) \left(
W_{13}+W_{31}\right) }
\nonumber\\
P_{3} &=&\frac{W_{12}W_{31}}{\left( W_{12}+W_{21}\right) \left(
W_{13}+W_{31}\right) }
\nonumber\\
P_{4} &=&\frac{W_{21}W_{31}}{\left( W_{12}+W_{21}\right) \left(
W_{13}+W_{31}\right) },
\label{eq:Pn}
\eea
where we have introduced the short notation
$W_{mn}= k_{n\rightarrow m}^{L}+k_{n\rightarrow m}^{R}$.
Since the qubits are of equal energy and $\left\vert S_{mn}^{L}\right\vert ^{2}=\left\vert
S_{mn}^{R}\right\vert ^{2}$, we have also utilized the fact that
$W_{13}=W_{24}$ and $W_{34}=W_{12}$ in deriving (\ref{eq:Pn}).

The rate constants depend on the particular choice of the system-bath interaction operator
and the bath Hamiltonian.
For example, assuming the reservoirs include a collection of harmonic modes
and that the bath operator coupled to the system is a displacement operator,
\bea
H_{\nu}=\sum_j\omega_jb_{\nu,j}^{\dagger}b_{\nu,j}, \,\,\,\,
B_{\nu}=\sum_j\lambda_{\nu,j}(b_{\nu,j}^{\dagger}+b_{\nu,j}),
\eea
the relaxation (excitation) rate with $m>n$ ($m<n)$) reduces to
\bea
k_{m\rightarrow n}^{\nu}&=&\Gamma_{B,\nu}(\omega_{mn})[n_B^{\nu}(\omega_{mn})+1],
\nonumber\\
k_{n\rightarrow m}^{\nu}&=&\Gamma_{B,\nu}(\omega_{mn})[n_B^{\nu}(\omega_{mn}),
\label{eq:bosonbath}
\eea
using the definition (\ref{eq:rate}). Here
$\Gamma_{B,\nu}(\omega)=2\pi\sum_j\lambda_{\nu,j}^2\delta(\omega-\omega_j)$
and $n_B^{\nu}(\omega)=[e^{\omega/T_{\nu}}-1]^{-1}$ is the Bose-Einstein distribution.
Another physical setup is the spin reservoir, including a collection of $P$ noninteracting spins,
\bea
H_{\nu}=\sum_{p=1}^{P} h_{\nu,p},\,\,\,\, B_{\nu}=\sum_{p=1}^Pb_{\nu,p}.
\eea
Each spin is described by the two eigenstates ($i=0,1$)
$|i\rangle_p$ and eigenenergies $\epsilon_p(i)$. In this case
the relaxation rate becomes
\bea
k_{m\rightarrow n}^{\nu}=\Gamma_{S,\nu}(\omega_{mn})n_S^{\nu}(-\omega_{mn})
\eea
with the spin occupation factor $n_S^{\nu}(\omega)=[e^{\omega/T_{\nu}}+1]^{-1}$
and the  effective spin-bath-system coupling $\Gamma_{S,\nu}(\omega )=2\pi \sum_{p}\left\vert
\left\langle 0\right\vert_{p}b_{\nu,p}\left\vert 1\right\rangle _{p}\right\vert ^{2}\delta (\omega
+\epsilon_{p}(0)- \epsilon_p(1))$. For details see Ref. \cite{WuRectif}.


\subsection{Heat current}

Formally, the expectation value of the current, calculated, e.g., at the left contact, is given by
$J_{L}=\frac{i}{2}$Tr$([H_{L}-H_{S},V_{L}]\rho )$,
$\rho$ is the total density matrix \cite{CurrentE}.
In steady state, the expectation value of the interaction is zero, Tr$(\frac{\partial V_{L}}{
\partial t}\rho )=i$Tr$([H_{L}+H_{S},V_{L}]\rho )$,
and this expression reduces to $J_{L}=i${\rm Tr}$([V_{L},H_{S}]\rho )$.
Using the system-diagonal representation,
it is straightforward to show that the steady state current becomes \cite{WuRectif}
\bea
J_{L}=i\sum_{m,n} \omega _{mn}S_{mn}^{L}{\rm Tr}_{B}(B^{L}\rho _{mn}),
\eea
where $\omega_{mn}=E_{m}-E_{n}$.
Under the Born-Markov approximation used above to resolve the population dynamics, a
second-order expression for the steady-state current can be further obtained \cite{WuRectif}
\bea
J_{L}=\frac{1}{2}\sum \omega _{mn}\left\vert S_{mn}^{L}\right\vert
^{2}P_{n}(k_{n\rightarrow m}^{L}-k_{n\rightarrow m}^{R}).
\eea
We apply this expression on the two-qubit model. A  somewhat tedious calculation shows that the
current is given by a sum of two terms
\bea
J_{L} &=&\frac{\omega _{21}(k_{1\rightarrow 2}^{L}k_{2\rightarrow
1}^{R}-k_{2\rightarrow 1}^{L}k_{1\rightarrow 2}^{R})}{2[k_{1\rightarrow
2}^{L}+k_{2\rightarrow 1}^{R}+k_{2\rightarrow 1}^{L}+k_{1\rightarrow 2}^{R}]}
\nonumber\\
&&+\frac{\omega _{31}(k_{1\rightarrow 3}^{L}k_{3\rightarrow
1}^{R}-k_{3\rightarrow 1}^{L}k_{1\rightarrow 3}^{R})}{2[k_{1\rightarrow
3}^{L}+k_{3\rightarrow 1}^{R}+k_{3\rightarrow 1}^{L}+k_{1\rightarrow 3}^{R}]}.
\label{eq:twocurrG}
\eea
%
\subsection{Concurrence}

The entanglement of formation is a monotonically
increasing function of the Wootters' concurrence \cite{Wooters}.
We calculate next the nonequilibrium concurrence in the two-qubit model.
The states $\left\vert 1\right\rangle $ and $\left\vert
4\right\rangle$ are entangled, and the nature of the entanglement may remain in
the final steady state, which is described by the diagonal reduced density
matrix $\rho _{d}=$ diag$(P_{1},P_{2},P_{3},P_{4})$, in the eigenbasis of the
system Hamiltonian $H_{S}$. On the other hand, in the uncoupled basis
$\left\vert \downarrow \downarrow \right\rangle ,\left\vert \downarrow
\uparrow \right\rangle ,\left\vert \uparrow \downarrow \right\rangle $ and $%
\left\vert \uparrow \uparrow \right\rangle $, the reduced density matrix
is given by a nondiagonal form,
\bea
\rho =\left(
\begin{array}{cccc}
P_{2} & 0 & 0 & 0 \\
0 & \frac{P_{1}+P_{4}}{2} & \frac{P_{4}-P_{1}}{2} & 0 \\
0 & \frac{P_{4}-P_{1}}{2} & \frac{P_{1}+P_{4}}{2} & 0 \\
0 & 0 & 0 & P_{3}
\end{array}
\right)
\eea
Using the general form given in, e.g., \cite{Fei11}, the
concurrence can be expressed in the uncoupled basis by,
\begin{equation}
C(T_{L},T_{R})=\max (2P_{\max }-P_{1}-P_{4}-2\sqrt{P_{2}P_{3}},0)
\label{concurrence}
\end{equation}
where $P_{\max }=\max (P_{1},P_{4},\sqrt{P_{2}P_{3}})$.
This function in general depends on the temperatures of both
thermal baths.

\begin{figure}
\hspace{2mm}
{\hbox{\epsfxsize=70mm \epsffile{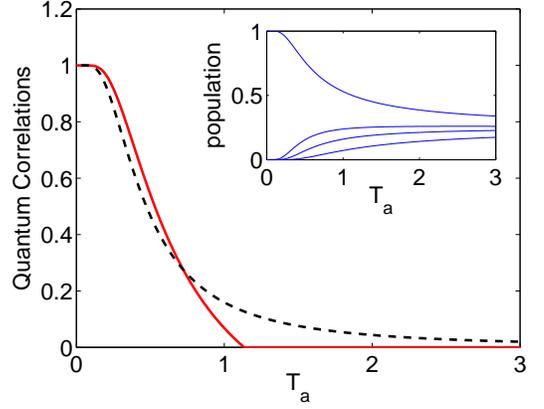}}} \caption{
Concurrence (full) and discord (dashed) in an equilibrium system, $T_a=T_L=T_R$,
The inset displays the levels population, top ($P_1$) to bottom ($P_4$).
The physical parameters are
$\epsilon=0.2$, $\kappa=1$. Bosonic reservoirs were adopted.}
\label{Fig1}
\end{figure}

\subsection{Discord}

Quantum discord \cite{Qcorr}, quantifying nonlocal correlations, is given by the
difference between the quantum mutual information $\mathcal{I}(\rho)$
and the classical correlation $\mathcal{C}(\rho)$,
\bea
\mathcal{Q}(\rho )=\mathcal{I}(\rho )-\mathcal{C}(\rho).
\eea
The analytical expressions for the classical correlation and the
quantum discord were obtained for a class of $X$ states
in Ref. \cite{Fei11}.
Adjusting these expression to our model, the quantum mutual information
can be written in terms of the steady state populations (\ref{eq:Pn}) as
\bea
&&\mathcal{I}(\rho) =
\nonumber\\
&&2-\log
_{2}[(1-P_{2}+P_{3})^{1-P_{2}+P_{3}}(1+P_{2}-P_{3})^{1+P_{2}-P_{3}}]
\nonumber\\
&&
+P_{1}\log _{2}P_{1}+P_{2}\log _{2}P_{2}+P_{3}\log _{2}P_{3}+P_{4}\log
_{2}P_{4},
\nonumber\\
\eea
whereas the classical correlation is given by
\bea
&&\mathcal{C}(\rho)=
\nonumber\\
&&1-\frac{1}{2}\log
_{2}[(1-P_{2}+P_{3})^{1-P_{2}+P_{3}}(1+P_{2}-P_{3})^{1+P_{2}-P_{3}}]
\nonumber\\
&&-\min
\{S_{1},S_{2}\}.
\eea
Here
\bea
&&S_{1} =
\nonumber\\
&&-P_{2}\log_{2}\left(\frac{2P_{2}}{1+P_{2}-P_{3}} \right)
-\left(\frac{P_{1}+P_{4}}{2}\right)\log_{2}\left(\frac{P_{1}+P_{4}}{
1+P_{2}-P_{3}}\right)
\nonumber\\
&&-\left(\frac{P_{1}+P_{4}}{2}\right)\log_{2}\left(\frac{P_{1}+P_{4}}{
1-P_{2}+P_{3}}\right)
-P_{3}\log_{2}\left(\frac{2P_{3}}{1-P_{2}+P_{3}}\right)
\nonumber\\
\eea
and
\bea
S_{2}=1-\frac{1}{2}\log _{2}[(1-K)^{1-K}(1+K)^{1+K}].
\eea
The coefficient $K$ is defined as
$K=\sqrt{(P_{2}-P_{3})^{2}+(P_{1}-P_{4})^{2}}$.


\begin{figure}
\hspace{2mm}
{\hbox{\epsfxsize=65mm \epsffile{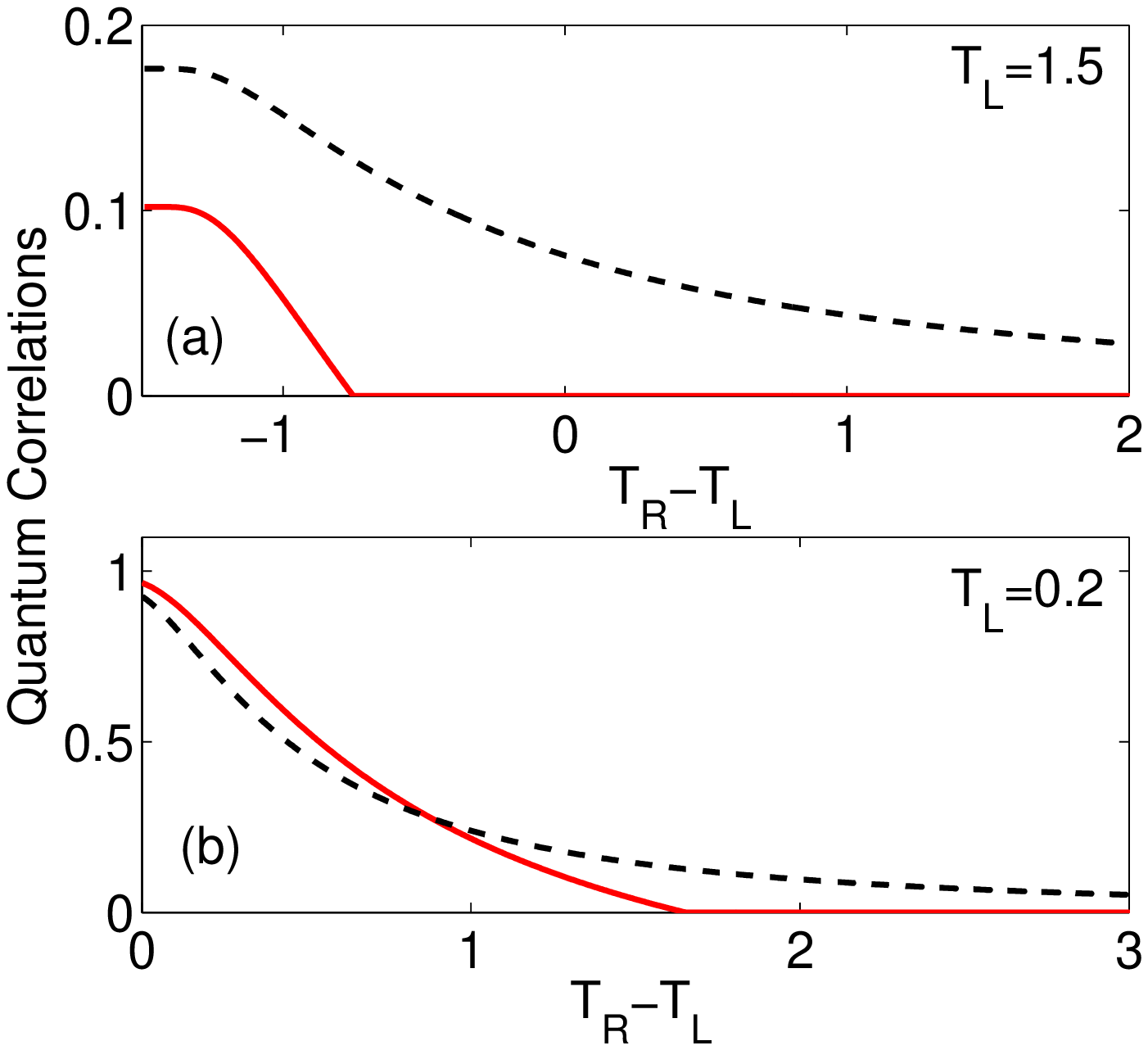}}} \caption{ (a)
Nonequilibrium thermal correlations for a pair of qubits coupled to boson baths.
Concurrence (full) and discord
(dashed) for $T_L=1.5$, $T_R$ is modified. (b) Same for $T_L=0.2$.
The two-qubit parameters are $\epsilon=0.2$, $\kappa=1$ and we set $\Gamma_{B,\nu}=1$.
}\label{Fig2}
\end{figure}

\begin{figure}
\hspace{2mm}
{\hbox{\epsfxsize=65mm \epsffile{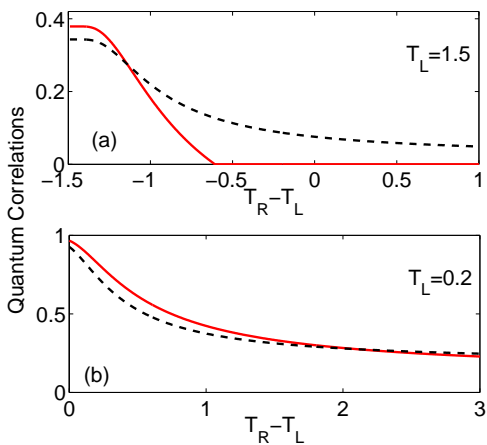}}} \caption{ (a)
Nonequilibrium thermal correlations for a pair of qubits coupled to spin baths.
Concurrence (full) and discord
(dashed) for $T_L=1.5$, $T_R$ is modified. (b) Same for $T_L=0.2$.
The two-qubit parameters are $\epsilon=1$ and  $\kappa=0.2$,  and we set $\Gamma_{S,\nu}=1$.
}\label{Fig3}
\end{figure}


\section{Examples}

In what follows we analyze the quantum correlations, concurrence and
discord, for various nonequilibrium conditions, showing that quantum
discord survives at relatively high temperatures where concurrence
is zero. We also study an asymmetric scenario demonstrating that
quantum correlations may exist even at large temperature biases. We
will typically use the following parameters: spin energy
$\epsilon=0.2$ and large spin-spin interaction $\kappa=1$. We will
also assume that the relaxation rates do not depend on energy, thus
treat them as constants, $\Gamma_{B,\nu}$ and $\Gamma_{S,\nu}$. The
factor $1/\sqrt2$ in Eq. (\ref{eq:S}) is absorbed into the
definition of the rates $\Gamma$.
%

We begin our analysis with an equilibrium situation, $T_a=T_L=T_R$, and compare the concurrence and
the discord measures at different temperatures. Fig. \ref{Fig1}
shows that at very low temperatures both measures yield the same
result. At higher temperatures (yet $T_a<\kappa$)
concurrence is slightly larger than discord. At even higher temperatures, $T_a>\kappa$,
 concurrence suddenly diminishes \cite{EntagD} while quantum discord is still finite, slowly
decreasing to zero. This is in accord with the fact
that discord quantifies non-classical correlations
beyond entanglement.
Generally, both quantum correlations decay with temperature due to thermal relaxation effects.
As can be inferred from the inset, when the ground state population falls below $\sim 1/2$ the concurrence dies.
We also found that these observations were not sensitive to the reservoirs properties, and similar trends
were obtained using either boson or spin baths.

Fig. \ref{Fig2} displays the nonequilibrium thermal correlations,
keeping $T_L$ fixed and changing $T_R$. In the classical limit, at
high temperatures (top panel) the discord overcomes the concurrence,
even  when $T_R$ is low. In contrast, at low temperatures (bottom panel) the opposite
trend is observed till a crossover value beyond which the
concurrence dies \cite{EntagD} yet the discord is finite.
While in Fig. \ref{Fig2} bosonic reservoirs have been adopted, Fig.
\ref{Fig3} displays the nonequilibrium thermal correlations
using spin baths. The main difference noted is that when $T_L$ is
quite high and $T_R$ is low [see Fig. \ref{Fig3}(a)], the spin-bath case
shows that concurrence is higher than discord, while the opposite
behavior is observed in Fig. \ref{Fig2}(a). This observation is in
accord with the notion that spin reservoirs are "more quantum" than
harmonic baths in the sense that harmonic modes can be represented
by two-level systems (spins) at low enough temperatures.

Next, we explore the role of spatial asymmetry on the survival of quantum correlations in
  (temperature) driven systems. Thermal rectification, an asymmetry of the
heat current for forward and reversed temperature gradients,
has been extensively analyzed in the last decade \cite{Rectifier,RectifierE,WuRectif}.
In a desirable rectifier the system behaves as an excellent heat conductor in one direction of the temperature bias,
while for the opposite direction it effectively acts as an insulator.
It is agreed that junctions incorporating anharmonic interactions with
some sort of spatial asymmetry should demonstrate this effect.
The two-qubit model, prepared with some asymmetry, e.g., assuming that the qubits asymmetrically couple
to, e.g., bosonic reservoirs, $\Gamma_{B,L} \neq \Gamma_{B,R}$, is expected to behave as a thermal
rectifier.

\begin{figure}
\vspace{8mm}
{\hbox{\epsfxsize=85mm \epsffile{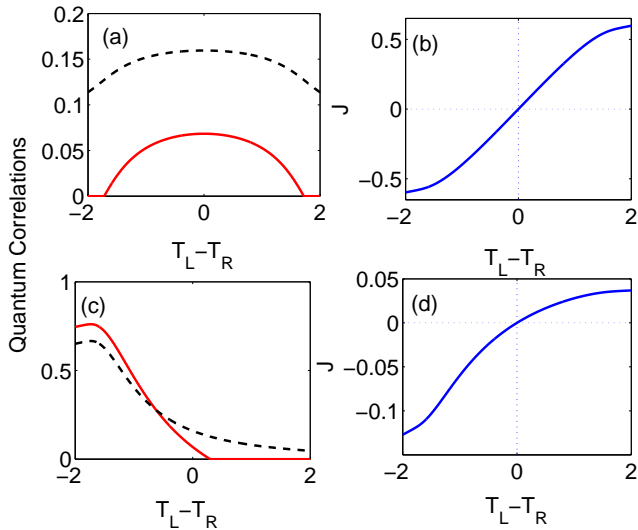}}} \caption{
Concurrence (full) and discord (dashed) in  symmetric (a) and asymmetric systems (c),
$T_a=1$ is fixed and the temperatures at the two ends are modulated,
$T_L=T_a+\Delta T$, $T_R=T_a-\Delta T$. In the symmetric case
$\Gamma_{B,L}=\Gamma_{B,R}=1$. Asymmetry is introduced by taking $\Gamma_{B,L}/\Gamma_{B,R}=20$
with $\Gamma_{B,L}=1$
Panels (b) and (d) further display the current for the symmetric and asymmetric cases, respectively.
In all panels $\epsilon=0.2$, $\kappa=1$.}
\label{Fig4}
\end{figure}

Figure \ref{Fig4} indeed shows the emergence of the thermal
rectifying effect upon turning on the asymmetry. The current in Fig.
\ref{Fig4}(b) is symmetric since $\Gamma_{B,L} = \Gamma_{B,R}$. In
contrast, in  Fig. \ref{Fig4}(d) the heat current is larger (in
magnitude) when the system is more strongly linked to the cold
reservoir ($T_L<T_R$ and $\Gamma_{B,L}>\Gamma_{B,R}$)
\cite{Rectifier}. Here the averaged temperature is $T_a=1$ with
$T_L=T_a+\Delta T$ and $T_R=T_a-\Delta T$. While the effect of
thermal rectification is well understood, here we demonstrate that
the transition between the fairly conducting phase ($\Delta T<0$)
and the poorly conducting phase ($\Delta T>0$) corresponds to a
turnover in the transport mechanism. In the symmetric setup discord
and concurrence are symmetric functions, see Fig. \ref{Fig4}(a),
$C(\Delta T)=C(-\Delta T)$ and $\mathcal Q(\Delta T)=\mathcal
Q(-\Delta T)$. Fig. \ref{Fig4}(c) shows a different behavior in the
presence of asymmetry: When the bias is negative quantum
correlations persist, However, for the opposite polarity the
concurrence is zero and discord is diminishing.

We can reason this behavior by noting that when the system is more strongly attached to the cold bath,
$\Gamma_{B,L}>\Gamma_{B,R}$ and $T_R>T_L$,
the ground state population is larger than that expected in the opposite case.
Since the ground state is an entangled (singlet) state,
$\left\vert 1\right\rangle =\frac{1}{\sqrt{2}}(\left\vert \downarrow
\uparrow \right\rangle -\left\vert \uparrow \downarrow \right\rangle$,
its large population reflects energy transmission process assisted
by a quantum correlated system state.

\begin{figure}
\vspace{8mm}
{\hbox{\epsfxsize=85mm \epsffile{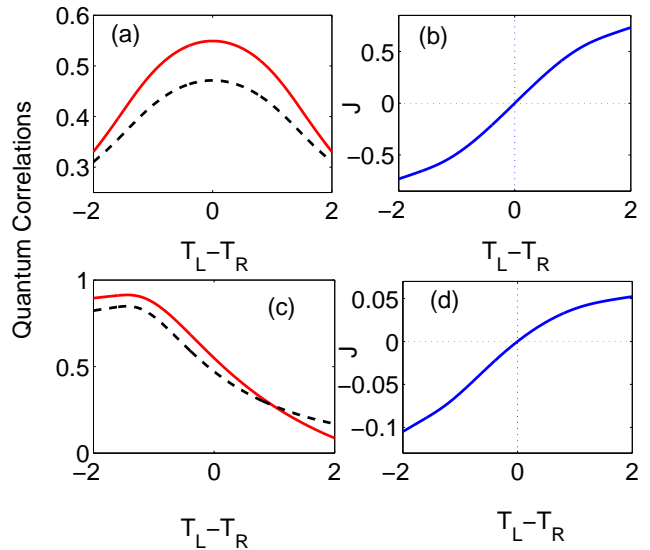}}} \caption{
Concurrence (full) and discord (dashed) in  symmetric (a)  and asymmetric systems (c).
Panels (b) and (d) display the current for the  symmetric and asymmetric situations, respectively.
Parameters are the same as in Fig. \ref{Fig4}, but the inter-spin interaction has been  increased to  $\kappa=2$.
}
\label{Fig5}
\end{figure}

By further increasing $\kappa$ one can extend the range over which quantum correlations survive,
see Fig. \ref{Fig5}.
For the asymmetric setup again we note that
when $T_L<T_R$ the current is large, in comparison to the opposite polarity,
and that entanglement measures are close to unity.
For the reversed case, $T_L>T_R$, the current magnitude is lowered, and quantum correlations
are being suppressed.
In particular, for $T_L-T_R=-2$, discord is large, $\mathcal Q\sim 1$, while at $T_L-T_R=2$ it is reduced
by an order of magnitude, $\mathcal Q\sim 0.1$.
Another interesting observation is that both discord and concurrence display a non-monotonic behavior at large negative bias,
reflected in a small maxima around $\Delta T\sim -1.7$ \cite{EntagT1,EntagT2,EntagT3}.

To conclude this section, by switching the sign of the temperature
bias one can control the magnitude of the heat current in asymmetric
spin chains and the underlying transport mechanism, as reflected by
the survival or suppression of quantum correlations in the system.

\vspace{8mm}

\section{Summary}

We detailed here a simple model of a many-body open quantum system
which could be analytically solved, useful for analyzing the role of
the temperature gradient on quantum correlations in a conducting
nanojunction. Our calculations manifest that for symmetric systems
under large temperature gradients and at high temperatures quantum
discord can be maintained, slowly decaying with $\Delta T$, whereas
concurrence typically dies.
In the presence of asymmetry we found that quantum correlations can survive in
one direction of the temperature gradient,
while they diminish when reversing the bias direction.

The present analysis could be generalized for describing transport
and quantum correlations in longer-linear spin chains. One could
also treat other systems, e.g., large spins, or adopt unequal
reservoirs at the two ends, for example, assuming the system is
coupled to both a solid and a metal \cite{WuRectif}. By
complementing transport studies with the calculation of quantum
correlations one can estimate and corroborate the role of quantum
effects in the transport process. This might be useful for building
quantum devices operating in noisy-thermal environments under
nonequilibrium conditions \cite{Baowenspin}.

\begin{acknowledgments}
L. -A. Wu has been supported by the Ikerbasque Foundation Start-up, the
CQIQC grant and the Spanish MEC (Project No. FIS2009-12773-C02-02).
D. S. Acknowledges support from the NSERC discovery grant.
\end{acknowledgments}



\begin{thebibliography}{9}

\bibitem{Dubi}
Y. Dubi and M. Di Ventra,
Rev. Mod. Phys. {\bf 83}, 131 (2011).

\bibitem{Rectifier}
M. Terraneo, M. Peyrard, and G. Casati, Phys. Rev. Lett. {\bf 88},
094302 (2002);
B. Li , L. Wang, and  G. Casati, Phys. Rev. Lett. {\bf 93}, 84301 (2004);
D. Segal and A. Nitzan, Phys. Rev. Lett. 94, 034301 (2005);
B. Hu, L. Yang, and Y. Zhang, Phys. Rev. Lett. {\bf 97}, 124302 (2006);
L. A. Wu and D. Segal, Phys. Rev. Lett. {\bf 102}, 095503 (2009).

\bibitem{BaowenL}
L. Wang and B. Li, Phys. Rev. Lett.  {\bf 99}, 177208 (2007).

\bibitem{BaowenM}
L. Wang and B. Li, Phys. Rev. Lett. {\bf 101}, 267203 (2008).

\bibitem{RectifierE}
C. W. Chang, D. Okawa, A. Majumdar, and A. Zettl, Science {\bf 314},
1121  (2006).

\bibitem{MemoryE}
R. Xie, {\it et al.},
Adv. Funct. Mater. (2011). XXX

\bibitem{Kirczenow}
L. G. C. Rego and G. Kirczenow,
Phys. Rev. Lett. {\bf 81}, 232 (1998).


\bibitem{NanoL}
I. Savic, N. Mingo, and D. A. Stewart, Phys. Rev. Lett. {\bf 101}, 165502 (2008);
D. A. Stewart, I. Savic, and  N. Mingo, Nano Lett. {\bf 9},81 (2008).

\bibitem{Green}
N. Mingo and L. Yang,
Phys. Rev. B {\bf 68}, 245406 (2003);
N. Mingo, Phys. Rev. B {\bf 74}, 125402 (2006).


\bibitem{Tu}
J.-S. Wang, J. Wang, and J. T. L\"u,
Euro. Phys. J. B {\bf 62}, 381 (2008).


\bibitem{Classical}
S. Lepri, R. Livi, and A. Politi,
Phys. Rep. {\bf 377}, 1 (2003);
A. Dhar, Adv. in Phys. {\bf 57}, 457 (2008).

\bibitem{BP}
P. Carruthers, Rev. Mod. Phys. {\bf 33}, 92 (1961).

\bibitem{Spohn}
H. Spohn, J. Stat. Phys. 124, 1041 (2006).

\bibitem{Mixed}
J.-S. Wang,
Phys. Rev. Lett. {\bf 99}, 160601 (2007).


\bibitem{Master}
H. Wichterich {\it et al.}, Phys. Rev. E {\bf 76}, 031115 (2007);
Y. Yan {\it et al.}, Phys. Rev. B {\bf 77}, 172411 (2008);

\bibitem{SegalM}
D. Segal, Phys. Rev. B {\bf 72}, 165426 (2006).

\bibitem{WuRectif}
L.-A. Wu, C. X. Yu, and D. Segal,
Phys. Rev. E {\bf 80}, 041103 (2009).

\bibitem{Qcorr}
H. Ollivier and W. H. Zurek, Phys. Rev. Lett. {\bf 88}, 017901 (2001);
W. H. Zurek, Rev. Mod. Phys. {\bf 75}, 715  (2003).

\bibitem{EntagT1}
L. Quiroga,  F. J. Rodríguez, M. E. Ramirez and R. Paris,
Phys. Rev. A {\bf 75}, 032308 (2007).

\bibitem{EntagT2}
I. Sinaysky, F. Petruccione, and D. Burgarth, Phys. Rev. A {\bf 78}, 062301 (2008).

\bibitem{EntagT3}
X. L. Huang, J. L. Guo , and X. X. Yi, Phys. Rev. A {\bf 80}, 054301 (2009).


\bibitem{Breuer}
H. P. Breuer and F. Petrucci, {\it The Theory of Open Quantum Systems}
(Oxford University Press 2002).

\bibitem{CurrentE}
L.-A. Wu and D. Segal, J. Phys. A: Math. and Theo. {\bf 42}, 025302 (2009).

\bibitem{Wooters}
W. K. Wootters, Phys. Rev. Lett. {\bf 80}, 2245 (1998).


\bibitem{Fei11}
B. Li, Z.-Z. Wang, and S.-M. Fei, Phys. Rev. A {\bf 83}, 022321 (2011).


\bibitem{EntagD}
J. H. Eberly and T. Yu.
Science {\bf 316}, 555 (2007);
M. P. Almeida, {\it et al.}, Science {\bf 316}, 579 (2007).

\bibitem{Baowenspin}
Y. Yan, C.-Q. Wu, and B. Li, Phys. Rev. B {\bf 79}, 014207 (2009).

\end{thebibliography}
\end{document}